\documentclass[letterpaper, 10 pt, conference]{ieeeconf}  

\IEEEoverridecommandlockouts                              

\usepackage{graphics} 
\usepackage{epsfig} 
\usepackage{mathptmx} 
\usepackage{times} 
\usepackage{amsmath} 
\usepackage{amssymb}  
\usepackage{booktabs}
\usepackage{color}
\usepackage{tikz}
\usepackage{placeins}
\usetikzlibrary{arrows.meta,calc}

\tikzset{
  wall/.style   ={line width=2.2pt, black, line cap=round},
  centre/.style ={line width=1.5pt, black, dash pattern=on 7pt off 5pt},
  flow/.style   ={line width=1.7pt, black, -{Stealth[length=9pt,width=7pt]}},
  lbl/.style    ={black, inner sep=1.5pt, font=\large},
  figcap/.style    ={black, font=\large, inner sep=0pt},
}

\newcommand{\corridor}[2]{%
  \begin{scope}[shift={(#1,0)}]
    \draw[flow] (4.0, 1.70)--(6.0, 1.70);  \node[lbl,anchor=west] at (6.15, 1.70) {$\rho_{#2,1,+}$};
    \draw[flow] (6.0, 0.65)--(4.0, 0.65);  \node[lbl,anchor=west] at (6.15, 0.65) {$\rho_{#2,1,-}$};
    \draw[flow] (4.0,-0.65)--(6.0,-0.65);  \node[lbl,anchor=west] at (6.15,-0.65) {$\rho_{#2,2,+}$};
    \draw[flow] (6.0,-1.70)--(4.0,-1.70);  \node[lbl,anchor=west] at (6.15,-1.70) {$\rho_{#2,2,-}$};
    \draw[flow] (1.30, 1.25)--(1.30,-1.25);
    \draw[flow] (0.80,-1.25)--(0.80, 1.25);
    \node[lbl,anchor=west] at (1.50, 1.00) {$\lambda^{\pm}_{#2,1\to 2}$};
    \node[lbl,anchor=west] at (1.50,-1.00) {$\lambda^{\pm}_{#2,2\to 1}$};
  \end{scope}%
}
\usepackage[T1]{fontenc}
\usepackage[utf8]{inputenc}
\newcommand{\R}{\mathbb R}
\newcommand{\C}{\mathbb C}
\newcommand{\cA}{\mathcal A}
\newcommand{\cB}{\mathcal B}
\newcommand{\cH}{\mathcal H}

\newcommand{\diag}{\operatorname{diag}}
\newcommand{\rank}{\operatorname{rank}}
\newcommand{\Ran}{\operatorname{Ran}}
\newcommand{\Ker}{\operatorname{Ker}}
\newcommand{\e}{\mathrm e}
\newcommand{\dd}{\,\mathrm d}

\newcommand{\sig}{\sigma}
\newcommand{\junction }{\mathcal J}
\newcommand{\col}{\operatorname{col}}
\newcommand{\totald}{R}
\newcommand{\decay}{d_c}
\newtheorem{theorem}{Theorem}[section]
\newtheorem{proposition}[theorem]{Proposition}
\newtheorem{lemma}[theorem]{Lemma}

\newtheorem{assumption}[theorem]{Assumption}

\title{\LARGE \bf
Localized Lane Guidance for a Bidirectional Pedestrian Cyclic Corridor: Equilibria and Generic Actuator Placement
}

\author{Adam Braun$^{1}$ and Nikolaos Bekiaris-Liberis$^{2}$
\thanks{*Funded by the European Union (ERC, C-NORA, 101088147). Views and opinions expressed are however those of the authors only and do not necessarily reflect those of the European Union or the European Research Council Executive Agency. Neither the European Union nor the granting authority can be held responsible for them.}
\thanks{$^{1}$Adam Braun is with Universit\'e Paris-Saclay, CNRS, CentraleSup\'elec, Laboratoire des signaux et syst\`emes, 91190, Gif-sur-Yvette, France. {\tt\small adam.braun@centralesupelec.fr}}%
\thanks{$^{2}$Nikolaos Bekiaris-Liberis with the Department of Electrical and Computer Engineering, Technical University of
Crete, University Campus, Chania, 73100, Greece
        {\tt\small bekiaris-liberis@ece.tuc.gr}}%
}

\begin{document}
\onecolumn

\maketitle
\thispagestyle{empty}
\pagestyle{empty}

\begin{abstract}
This paper studies bidirectional pedestrian flow on a cyclic network of corridors. Each corridor consists of two lanes, while pedestrians may enter or leave the network at junctions. The dynamics are based on a 1-D, bidirectional, two-lane Lighthill-Whitham-Richards-type model, in which lane changes are described through explicit, density-dependent transfer rates that favor motion toward the less congested lane. We propose a localized lane-guidance control input that biases lane-changing behavior without creating pedestrians or reversing their walking direction. Such guidance may represent, for example, recommendations provided by staff or information screens. Under meaningful assumptions on the external inflows, we prove the existence and uniqueness of an admissible uniform equilibrium. We then linearize the system around this equilibrium and study the placement of localized guidance zones through a spectral detectability condition. We prove that, for any prescribed decay rate, two guidance patches located in a single, suitably chosen corridor are sufficient to satisfy the spectral detectability condition for all critical adjoint eigenmodes. This condition is necessary for the exponential stabilizability of the linearized system. Numerical experiments illustrate the spectral properties of the model and study the robustness of the modal observability condition with respect to the geometry and distribution of the guidance patches.
\end{abstract}


\section{Introduction}
Dynamic modeling of pedestrian flow is pertinent to a broad range of
applications including, the study of epidemic transmission in dense
environments~\cite{abdulsalam2023coupling,haghani2022crowd,delis2026numerical}, emergency evacuation and safety
management~\cite{chen2021pedestrian,haghani2020optimising,twarogowska2014macroscopic, helbing2002simulation}, and the design and
operation of crowded facilities such as airport terminals, stadiums, and
public buildings~\cite{vermuyten2016review}. At the macroscopic scale, Partial
Differential Equations (PDEs) constitute a natural framework for describing pedestrian
densities and fluxes~\cite{hughes2002continuum,cristiani2014multiscale}. Bidirectional pedestrian motion in corridors has been studied both experimentally
and numerically, in particular with data-driven macroscopic models able to reproduce
counter-flow dynamics and assess segregation strategies
~\cite{appert2011two,motsch2018modeling,jiang2012numerical}.

Compared with the extensive literature devoted to modeling and simulation, strategies that actively modify pedestrian behavior are less
developed. In particular, interventional approaches constitute a relatively
underexplored direction in crowd management~\cite{haghani2020optimising}.
Existing approaches include dynamic marking, routing or speed recommendations, and directional segregation strategies designed to reduce congestion~\cite{delis2026numerical,motsch2018modeling,chu2017variable,li2022dynamic, kachroo2008pedestrian}. Such interventions are particularly relevant when modifying the infrastructure is not possible. They could be implemented through information screens, floor markings, or instructions provided by staff.

For infinite-dimensional linear systems, a classical necessary condition for stabilizability is the Fattorini-Hautus (FH) spectral condition~\cite{fattorini1966some}. Its verification is, however, strongly model-dependent. This abstract rank condition should be related to structural assumptions on the underlying PDE and the control action. Such reductions have been obtained for several classes of PDEs, for instance, in parabolic systems~\cite{badra2014fattorini,boyer2014hautus}, or first-order hyperbolic systems~\cite{coron2016stabilization}. Our objective here is precisely to verify the (FH) spectral condition directly from explicit and physically interpretable
assumptions on a pedestrian-flow model and on the placement of localized
guidance zones.

We adapt the two-way multi-lane model (ML-CAR) of~\cite{appert2011two} to a cyclic
corridor with density-dependent lane transfers and localized
guidance. We derive physically interpretable conditions ensuring an admissible
free-flow equilibrium and show that, under a non-degeneracy condition on the
lane pressures, two guidance patches in one corridor detect all critical adjoint
eigenmodes. Moreover, we establish that the patch geometries satisfying the (FH) condition form a residual set
of full Lebesgue measure. Numerical experiments illustrate the spectral and
conditioning properties.


The remainder of the paper is organized as follows.
Section~\ref{sec:model} introduces the nonlinear pedestrian flow model and its
equilibria. Section~\ref{sec:linear} derives the linearized dynamics and
formulates the spectral detectability problem. Section~\ref{sec:main_result}
contains the main result, and
Section~\ref{section:numerics} provides a numerical illustration. Concluding remarks are provided in Section~\ref{sec:conclusion}.

\paragraph*{Notation}
We use $\C$ and $\R$ for the complex and real fields, $I$ for the identity
operator, and $\mathbf 1_\omega$ for the indicator function of a measurable set
$\omega$.
\section{Pedestrian Model}
\label{sec:model}
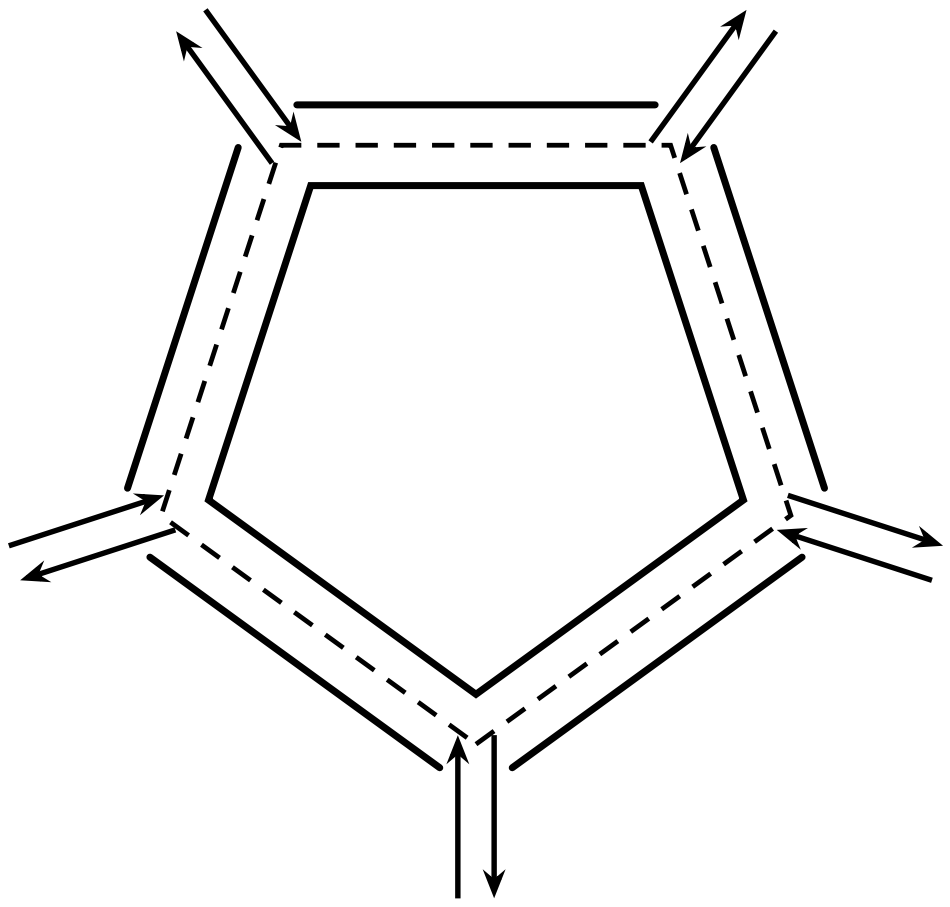
\begin{figure}
    \centering
\scalebox{0.6}{\begin{tikzpicture}[x=1cm,y=1cm]
  \def\Ro{4.2}\def\Ri{3.1}\def\Rm{3.65}
  \foreach \i in {0,...,4}{
    \pgfmathsetmacro{\a}{-90+72*\i}
    \coordinate (O\i) at (\a:\Ro);
    \coordinate (I\i) at (\a:\Ri);
    \coordinate (M\i) at (\a:\Rm);
  }
  \foreach \i/\j in {0/1,1/2,2/3,3/4,4/0}{
    \draw[wall] ($(O\i)!0.10!(O\j)$) -- ($(O\i)!0.90!(O\j)$);
  }
  \draw[wall] (I0)--(I1)--(I2)--(I3)--(I4)--cycle;
  \draw[centre] (M0)--(M1)--(M2)--(M3)--(M4)--cycle;
  \foreach \i in {0,...,4}{
    \pgfmathsetmacro{\a}{-90+72*\i}
    \draw[flow] ($(\a:3.55)+(\a+90:0.20)$) -- ($(\a:5.35)+(\a+90:0.20)$);
    \draw[flow] ($(\a:5.35)+(\a+90:-0.20)$) -- ($(\a:3.55)+(\a+90:-0.20)$);
  }
\end{tikzpicture}}
\caption{Example of a cyclic network with $N=5$ corridors.}
    \label{fig:pentagone}
\end{figure}

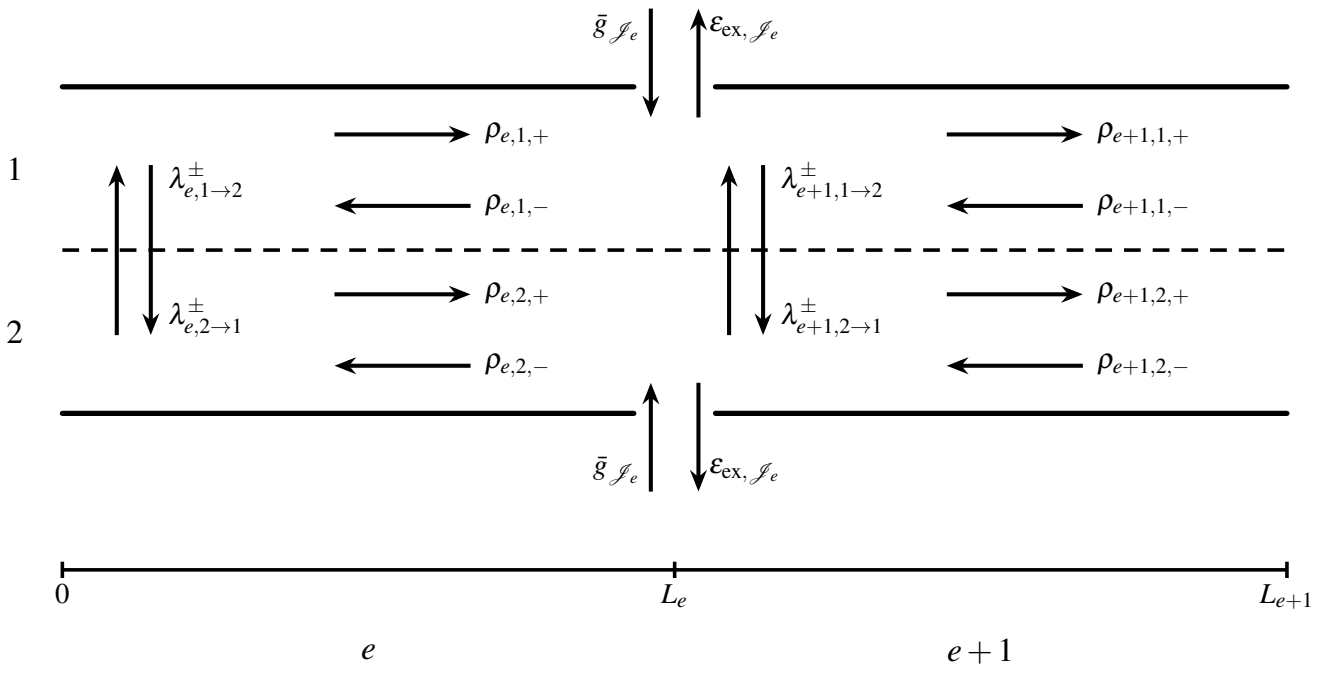
\begin{figure}
    \centering
    \scalebox{0.45}{\begin{tikzpicture}[x=1cm,y=1cm]
  \draw[wall] (0,2.4)--(8.4,2.4);   \draw[wall] (9.6,2.4)--(18,2.4);
  \draw[wall] (0,-2.4)--(8.4,-2.4); \draw[wall] (9.6,-2.4)--(18,-2.4);
  \draw[centre] (0,0)--(18,0);

  \corridor{0}{e}
  \corridor{9}{e+1}

  \node[font=\Large,black] at (-0.7, 1.20) {$1$};
  \node[font=\Large,black] at (-0.7,-1.20) {$2$};

  \draw[flow] (8.65,3.55)--(8.65,1.95);
  \draw[flow] (9.35,1.95)--(9.35,3.55);
  \node[lbl,anchor=east] at (8.55,3.25) {$\bar g_{\mathcal{J}_e}$};
  \node[lbl,anchor=west] at (9.45,3.25) {$\epsilon_{{\rm ex},\mathcal{J}_e}$};
  \draw[flow] (8.65,-3.55)--(8.65,-1.95);
  \draw[flow] (9.35,-1.95)--(9.35,-3.55);
  \node[lbl,anchor=east] at (8.55,-3.25) {$\bar g_{\mathcal{J}_e}$};
  \node[lbl,anchor=west] at (9.45,-3.25) {$\epsilon_{{\rm ex},\mathcal{J}_e}$};

  \draw[line width=1.5pt,black] (0,-4.7)--(18,-4.7);
  \foreach \x in {0,9,18}{\draw[line width=1.5pt,black] (\x,-4.58)--(\x,-4.82);}
  \node[lbl,anchor=north] at (0,-4.85)  {$0$};
  \node[lbl,anchor=north] at (9,-4.85)  {$L_e$};
  \node[lbl,anchor=north] at (18,-4.85) {$L_{e+1}$};
  \node[lbl,font=\Large] at (4.5,-5.9)  {$e$};
  \node[lbl,font=\Large] at (13.5,-5.9) {$e+1$};

\end{tikzpicture}}
    \caption{Representation of two consecutive corridors.}
    \label{fig:corridor}
\end{figure}

We consider a network of $N\ge2$ corridors connected in a cycle, as
illustrated in Fig.~\ref{fig:pentagone}.
Pedestrians enter and leave the cyclic corridor at the junctions. Each
corridor follows the ML-CAR model of~\cite{appert2011two} (a particular Lighthill-Whitham-Richards-type model), which consists of two
lanes with bidirectional motion and lane transfers, as illustrated in
Fig.~\ref{fig:corridor}.
The lane transfer rates
specified below are a modeling choice. Control is implemented through localized guidance zones that influence lane-changing behavior without directly modifying the total pedestrian flow (for example, with agents providing recommendations to pedestrians). 
The objective of studying stabilizability depending on the number/location of guidance zones, is formulated for the linearized dynamics around a stationary equilibrium point.
All physical variables can be found in Table~\ref{tab:physical-variables}.
\begin{table}[t]
\centering
\caption{Variables and parameters of the model}
\label{tab:physical-variables}
\begin{tabular}{ll}
\toprule
Symbol & Physical interpretation \\
\midrule
$\rho_{e,k,\alpha}(t,x)$
& Pedestrian density in corridor $e$, lane $k$, direction $\alpha$
\\

$\totald_{e,k}$
& Total density in lane $k$,
$\totald_{e,k}=\rho_{e,k,+}+\rho_{e,k,-}$
\\

$\rho^\star$
& Limiting admissible lane density
\\

$V$
& Desired free walking speed
\\

$M_{e,k}$
& Coefficient of the linear pressure law
\\

$P_{e,k}$
& Lane pressure,
$P_{e,k}=M_{e,k}\totald_{e,k}$


\\

$f_{e,k,\alpha}$
& Physical pedestrian flux,
$f_{e,k,\pm}=\rho_{e,k,\alpha}(\pm V\mp P_{e,k})$
\\

$\lambda_{e,i\to j}^{\alpha}$
& Lane-changing rate from lane $i$ to lane $j$, direction $\alpha$
\\

$\lambda_0$
& Reference lane-changing frequency
\\

$\theta$
& Sensitivity variable for lane-choice response
 \\

$\eta_{e,\alpha}(t,x)$
& Guidance bias acting on lane-changing behavior
 \\

$\varepsilon_{{\rm ex},J}$
& Fraction of flow leaving the concourse at junction $J$
 \\

$p_J$
& Fraction of flow continuing at $J$,
$p_J=1-\varepsilon_{{\rm ex},J}$
 \\
$\bar g_J$
& Constant external pedestrian arrival flux at junction $J$
\\
\bottomrule
\end{tabular}
\end{table}
\subsection{Pedestrian Dynamics}
Let $\mathcal E=\{1,\ldots,N\}$, with indices modulo $N$. Corridor $e$ is represented by the spacial variable $x\in(0,L_e)$, $L_e>0$.
For lane $k\in\{1,2\}$ and desired direction $\alpha\in\{+,-\}$,
$\rho_{e,k,\alpha}$ denotes density, set
\begin{equation}
 \rho_e=\operatorname{col}(\rho_{e,1,+},\rho_{e,1,-},
                  \rho_{e,2,+},\rho_{e,2,-}),
 ~ \totald_{e,k}=\rho_{e,k,+}+\rho_{e,k,-}.
 \label{eq:qe-def}
\end{equation}
With $V>0$ and $M_{e,k}>0$, the pressure and signed fluxes are
\begin{equation}
 \left\{\begin{aligned}
 &P_{e,k}=M_{e,k}\totald_{e,k},\\
 &f_{e,k,+}=\rho_{e,k,+}(V-P_{e,k}),\\
 &f_{e,k,-}=\rho_{e,k,-}(-V+P_{e,k}).
 \end{aligned}\right.
 \label{eq:fluxes}
\end{equation}
This is a special case  from~\cite[Section 4.1.1]{appert2011two} and constitutes a Greenshield fundamental diagram~\cite{greenshields1935study},
set
\begin{equation}F_e=\operatorname{col}(f_{e,1,+},f_{e,1,-},f_{e,2,+},f_{e,2,-}).\end{equation}
The balance laws, with given initial data $\rho_e^0$, are\footnote{The present preliminary work does not address the well-posedness of the full nonlinear network system, as it focuses on the properties of the linearized system.}
\begin{equation}
 \left\{\begin{aligned}
 &\partial_t \rho_e+\partial_x F_e(\rho_e)=S_e(\rho_e,\eta_e),
       &&t>0,\quad 0<x<L_e,\\
 &\rho_e(0,x)=\rho_e^0(x).
 \end{aligned}\right.
 \label{eq:nonlinear-corridor-system}
\end{equation}
For a maximum admissible lane density $\rho^\star>0$,  define
\begin{equation}
 \chi(\rho)=1-\rho/\rho^\star~(0\le \rho<\rho^\star),
 \quad \sig(d)=\frac1{1+\e^{-d}}.
 \label{eq:chi}
\end{equation}
We tacitly assume that
\begin{equation}
    \label{eq:max_density}
    \frac{V}{2M_{e,k}}<\rho^\star,
    \quad \forall e\in\mathcal E,~\forall k\in\{1,2\},
\end{equation}
which typically holds, as the quantity $\frac{V}{2M_{e,k}}$ represents the critical density of Greenshields' fundamental diagram corresponding to total flow (i.e., $f_{e,k,+} - f_{e,k,-}$) at a lane.
For $\lambda_0>0$ a reference lane-changing rate, $\theta>0$
a sensitivity scaling factor and $\eta_{e, \alpha}$ a guidance input, set
\begin{equation}
 \left\{\begin{aligned}
 &\lambda_{e,1\to2}^{\alpha}
 =\lambda_0\chi(\totald_{e,2})
 \sig\bigl((P_{e,1}-P_{e,2}+\eta_{e,\alpha})/\theta\bigr),\\
 &\lambda_{e,2\to1}^{\alpha}
 =\lambda_0\chi(\totald_{e,1})
 \sig\bigl((P_{e,2}-P_{e,1}-\eta_{e,\alpha})/\theta\bigr),\\
 &S_{e,1,\alpha}=\lambda_{e,2\to1}^{\alpha}\rho_{e,2,\alpha}
                  -\lambda_{e,1\to2}^{\alpha}\rho_{e,1,\alpha},
 \quad S_{e,2,\alpha}=-S_{e,1,\alpha},\\
 &S_e(\rho_e, \eta_e)=\col(S_{e,1,+}, S_{e,1,-}, S_{e,2,+}, S_{e,2,-}).
 \end{aligned}\right.
 \label{eq:rates}
\end{equation}
As in~\cite{appert2011two}, $S_e$ is conservative for each walking direction.
The rates favor transfers toward lower-pressure lanes, while $\chi$ removes
transfers toward lanes approaching $\rho^\star$. The input $\eta_{e,\alpha}$
hence represents guidance biasing lane choice only.
\subsection{Guidance Zones and Junctions}
A guidance zone incorporates two independent signals (control inputs), one for each walking
direction. For $N_e\geq 0$ patches of length $\omega_{e,j}\subset(0,L_e)$, write for $\alpha \in \{+,-\}$
\begin{equation}
 \eta_{e,\alpha}(t,x)=\bar\eta_{e,\alpha} +\sum_{j=1}^{N_e}
       \mathbf1_{\omega_{e,j}}(x)u_{e,j,\alpha}(t).
 \label{eq:indicator-control}
\end{equation}
These signals only manipulate lane choice, while they do not create pedestrians or reverse
walking direction, since $S_{e,1,\alpha}+S_{e,2,\alpha}=0$.
The nominal guidance input $\bar \eta_e =\col(\bar \eta_{e,+}, \bar \eta_{e, -})$ can be interpreted as fixed direction markings already present in the corridor, whereas $u_{e, j}$ is interpreted as the effect of instructions/guidance provided by localized staff or screens in real time. 
At the junction $\junction _e$ between $e$ and $e+1$, 
\begin{equation}
\begin{bmatrix}
 f_{e+1,1,+}(t,0)\\
 f_{e+1,2,+}(t,0)\\
 -f_{e,1,-}(t,L_e)\\
 -f_{e,2,-}(t,L_e)
\end{bmatrix}
=
p_{\junction_e}
\begin{bmatrix}
 f_{e,1,+}(t,L_e)\\
 f_{e,2,+}(t,L_e)\\
 -f_{e+1,1,-}(t,0)\\
 -f_{e+1,2,-}(t,0)
\end{bmatrix}
+
\bar g_{\junction_e},
\label{eq:junction-boundary}
\end{equation}
with
\begin{equation}
p_{\junction_e}
=
1-\varepsilon_{{\rm ex},\junction_e}
\in(0,1],
\quad
\varepsilon_{{\rm ex},\junction_e}\in[0,1),
\quad
\bar g_{\junction_e}\in\R_+^4.
\end{equation}
Boundary conditions~\eqref{eq:junction-boundary} imply that pedestrians retain their lane and direction. At junctions, the coefficient
$\varepsilon_{{\rm ex},\junction _e}$ indicates the percentage of pedestrians that leaves the concourse, and $\bar g_{\junction _e}$
is a fixed arrival flow rate. Hence, lane redistribution occurs only within the corridor (i.e., inside the domain).
\subsection{Equilibria and Assumptions}
Let $\bar g = (\bar g_{\junction_e})_{e\in \mathcal E}$ be a given external pedestrian arrival flux.
A point $\bar \rho =(\bar \rho_e)_{e\in \mathcal E}$ with associated fixed fluxes $\bar f_e$ is an equilibrium if there exists a nominal guidance $\bar \eta = (\bar \eta_e)_{e\in \mathcal E}$ such that for all $e\in \mathcal E$, $S_e(\bar \rho_e,\bar\eta_e)=0$, i.e., for all $\alpha\in\{+,-\}$,
\begin{equation}\label{eq:source_zero}
\bar\lambda^\alpha_{e,2\to1}
    \bar\rho_{e,2,\alpha}
    =
    \bar\lambda^\alpha_{e,1\to2}
    \bar\rho_{e,1,\alpha},
\end{equation}
and 
\begin{equation}
    \label{eq:boundary_equi}
    \begin{aligned}
    \col(
&\bar f_{e+1,1,+},
 \bar f_{e+1,2,+},
 -\bar f_{e,1,-},
 -\bar f_{e,2,-})
\\&=
p_{\junction_e}
\col(
 \bar f_{e,1,+},
\bar  f_{e,2,+},
 -\bar f_{e+1,1,-},
 -\bar f_{e+1,2,-}) + \bar g_{\junction_e}.
 \end{aligned}
\end{equation}
We next give conditions on the given external arrival flux
$\bar g$ and parameters of the system under which an admissible equilibrium is uniquely determined. Set
\begin{equation}
    P_{\rm cyc}
    :=
    \prod_{e=1}^{N}p_{\junction_e},
\end{equation}
by definition $P_{\rm cyc}>0$.
For all $e\in \mathcal E$ and all $k\in\{1,2\}$, define the positive and negative direction arrivals
\begin{align}
    \Gamma_{e,k}^{+}
    &:=
    \sum_{r=0}^{N-1}
    \left(
        \prod_{q=1}^{r}p_{\junction_{e-q}}
    \right)
    \bar g_{\junction_{e-1-r},k,+},
    \label{eq:Gamma-plus}
    \\
    \Gamma_{e,k}^{-}
    &:=
    \sum_{r=0}^{N-1}
    \left(
        \prod_{q=0}^{r-1}p_{\junction_{e+q}}
    \right)
    \bar g_{\junction_{e+r},k,-},
    \label{eq:Gamma-minus}
\end{align}
where an empty product is equal to one.  These quantities collect the
external arrivals which reach corridor $e$ after propagation through the
successive junctions.
We assume the following.
\begin{assumption}
\label{ass:model}
For all $e\in \mathcal E$ and all $k\in \{1,2\}$,
\begin{equation}
    \Gamma_{e,k}^{+}>0,
    \quad
    \Gamma_{e,k}^{-}>0,
    \label{eq:positive-effective-arrivals}
\end{equation}
and
\begin{equation}
0<
\frac{\Gamma_{e,k}^{+}+\Gamma_{e,k}^{-}}
{1-P_{\rm cyc}}
<
\frac{V^2}{4M_{e,k}}.
    \label{eq:inflow-admissibility}
\end{equation}
Moreover, assume
\begin{equation}
    P_{\rm cyc}<1,
    \label{eq:equilibrium-cycle-loss}
\end{equation}
i.e., there is at least one exit with $\varepsilon_{{\rm ex},\junction_e}>0$.
\end{assumption}
\begin{lemma}\label{lem:inflow_gives_equi}
    Let $\bar g$ be an arrival flux satisfying Assumption~\ref{ass:model}.
    Then, $\bar g$ uniquely determines an equilibrium point $\bar \rho$ and a nominal guidance $\bar \eta$ such that~\eqref{eq:source_zero},~\eqref{eq:boundary_equi} hold. 
\end{lemma}
The proof of Lemma~\ref{lem:inflow_gives_equi} is postponed to Appendix~\ref{appendix:proof_lem_{e+1}nflow_equi}.

From the proof of Lemma~\ref{lem:inflow_gives_equi} in Appendix~\ref{appendix:proof_lem_{e+1}nflow_equi}, it follows that~\eqref{eq:inflow-admissibility} guarantees that the reconstructed
equilibrium satisfies
\begin{equation}\label{eq:free_flow_regime}
0<
\bar{\totald}_{e,k}
<
\frac{V}{2M_{e,k}},
\end{equation}
and hence lies in the free-flow regime.\footnote{Extensions to congested or mixed
regimes would require additional assumptions to ensure hyperbolicity and
boundary compatibility of the system. Thus, they are not studied within this preliminary work.}
Moreover, the proof of Lemma~\ref{lem:inflow_gives_equi} in Appendix~\ref{appendix:proof_lem_{e+1}nflow_equi}, implies
\begin{equation}
\frac{\bar\rho_{e,k,+}}{\bar\rho_{e,k,-}}
=
\frac{\Gamma^+_{e,k}}{\Gamma^-_{e,k}}.
\end{equation}
Hence, if $
\Gamma^+_{e,1}>\Gamma^-_{e,1}
$ and $
\Gamma^-_{e,2}>\Gamma^+_{e,2}$,
then the corresponding equilibrium exhibits directional segregation, with lane~$1$ mostly occupied by pedestrians moving in the positive direction and lane~$2$ mostly occupied by pedestrians moving in the negative direction. The stability of such an equilibrium would ensure smoother traffic in the cylic corridor (see Section 3.4 in~\cite{motsch2018modeling}).
\begin{assumption}
    \label{ass:transport-nondegeneracy}
    There exists $e_\star\in \mathcal E$, such that
    \begin{equation}\label{eq:non_degen_trans}
    M_{e_\star,1}
    \bigl(
        \Gamma_{e_\star,1}^{+}
        +
        \Gamma_{e_\star,1}^{-}
    \bigr)
    \neq
    M_{e_\star,2}
    \bigl(
        \Gamma_{e_\star,2}^{+}
        +
        \Gamma_{e_\star,2}^{-}
    \bigr).
\end{equation}
From the proof of Lemma~\ref{lem:inflow_gives_equi} in Appendix~\ref{appendix:proof_lem_{e+1}nflow_equi},~\eqref{eq:non_degen_trans} is equivalent to 
 \begin{equation} \label{eq:cond_corridor_non_degen}
    M_{e_\star,1}\bar{\totald}_{e_\star,1}
    \neq
    M_{e_\star,2}\bar{\totald}_{e_\star,2}.
\end{equation}
\end{assumption}
In turn, condition~\eqref{eq:cond_corridor_non_degen} implies that, at equilibrium, there exists at least one corridor link at which the pressure in one lane is different than the pressure in the other, so that a bias in density at one lane of one direction over the other can be achieved via the nominal guidance. Such a bias may be beneficial as it may reduce friction in pedestrians' movements, which in turn may be beneficial for pedestrians' comfort and travel time.

\section{Linearized System}
\label{sec:linear}
\subsection{Linearized Dynamic}
Let $\bar g = (\bar g_{\junction_e})_{e\in \mathcal E}$ an arrival flux satisfying Assumptions~\ref{ass:model},~\ref{ass:transport-nondegeneracy}. By Lemma~\ref{lem:inflow_gives_equi}, let $\bar \rho =(\bar \rho_e)_{e\in \mathcal E}$ and $\bar \eta = (\bar \eta_e)_{e\in \mathcal E}$ denote the associated equilibrium point and nominal guidance.
For all $e\in \mathcal E$, set
$
    r_e=\rho_e-\bar \rho_e.$
The linearization of~\eqref{eq:nonlinear-corridor-system} around
$(\bar \rho_e,\bar\eta_e)$ is
\begin{equation}
    \partial_t r_e+A_e\partial_xr_e
    =
    J_er_e+
    \sum_{j=1}^{N_e}
    \mathbf1_{\omega_{e,j}}B_eu_{e,j},
    \label{eq:linear-physical}
\end{equation}
where
\begin{equation}
    A_e=DF_e(\bar \rho_e),
    \quad
    J_e=D_qS_e(\bar \rho_e,\bar\eta_e),
    \quad
    B_e=D_\eta S_e(\bar \rho_e,\bar\eta_e).
    \label{eq:linear-matrices}
\end{equation}
In particular, we have
\begin{equation}
    A_e=\diag(A_{e,1},A_{e,2}),
    \label{eq:Ae-subsystem}
\end{equation}
where
\small
\begin{equation}
 A_{e,k}
 =
 \begin{bmatrix}
 V-M_{e,k}(2\bar\rho_{e,k,+}+\bar\rho_{e,k,-})
 &-M_{e,k}\bar\rho_{e,k,+}\\[1mm]
 M_{e,k}\bar\rho_{e,k,-}
 &-V+M_{e,k}(\bar\rho_{e,k,+}+2\bar\rho_{e,k,-})
 \end{bmatrix}.
 \label{eq:Aek}
\end{equation}
\normalsize
Under Assumption~\ref{ass:model}, we have~\eqref{eq:free_flow_regime}, hence
\begin{equation}
    \det A_{e,k}
    =
    -\bigl(V-M_{e,k}\bar{\totald}_{e,k}\bigr)
      \bigl(V-2M_{e,k}\bar{\totald}_{e,k}\bigr)
    <0.
    \label{eq:det-Aek}
\end{equation}
Moreover, for all $\alpha\in\{+,-\}$, set
\begin{equation}
    \bar d_{e,\alpha}
    :=
    \frac{
        M_{e,1}\bar{\totald}_{e,1}
        -
        M_{e,2}\bar{\totald}_{e,2}
        +
        \bar\eta_{e,\alpha}
    }{\theta}.
    \label{eq:dbar-alpha-linear}
\end{equation}
We have
\begin{equation}
    B_e
    =
    \begin{bmatrix}
        -\kappa_{e,+}&0\\
        0&-\kappa_{e,-}\\
        \kappa_{e,+}&0\\
        0&\kappa_{e,-}
    \end{bmatrix},
    \label{eq:Be}
\end{equation}
where
\begin{equation}
    \kappa_{e,\alpha}
    =
    \frac{\lambda_0\sig'(\bar d_{e,\alpha})}{\theta}
    \left[
        \chi(\bar{\totald}_{e,1})\bar\rho_{e,2,\alpha}
        +
        \chi(\bar{\totald}_{e,2})\bar\rho_{e,1,\alpha}
    \right]>0 .
    \label{eq:kappa-nonzero-guidance}
\end{equation}The explicit entries of the source $J_e$ are not required in the
stabilizability analysis.  What is essential is its
structure.  Because of the structure of $S_e(\bar \rho_e, \bar \eta_e)$ in~\eqref{eq:rates}
one has
\begin{equation}
    \operatorname{Ran}J_e
    \subset
    \mathcal D
    :=
    \left\{
        \begin{pmatrix}-z\\ z\end{pmatrix}
        :z\in\mathbb C^2
    \right\}.
    \label{eq:range-Je}
\end{equation}
The linearized flux perturbation in corridor $e\in \mathcal E$ is $A_er_e$. Since the
external arrivals are fixed, their perturbation vanishes.  Hence, at
the junction $\junction_e$ connecting corridor $e$ to $e+1$, the
linearization of~\eqref{eq:junction-boundary} is
\begin{equation}
\left\{
\begin{aligned}
    (A_{e+1}r_{e+1}(0))_j
    &=p_{\junction_e}(A_er_e(L_e))_j,
    &&j\in\{1,3\},\\
    (A_er_e(L_e))_i
    &=p_{\junction_e}(A_{e+1}r_{e+1}(0))_i,
    &&i\in\{2,4\}.
\end{aligned}
\right.
\label{eq:linear-boundary}
\end{equation}
The indices $1,3$ correspond to the positive walking direction and
$2,4$ to the negative walking direction.
\subsection{Abstract Formulation}

For the spectral analysis, we complexify the system and set
\begin{equation}
    \cH
    :=
    \prod_{e=1}^N L^2(0,L_e;\C^4),
    \quad
    \langle r,z\rangle_{\cH}
    :=
    \sum_{e=1}^N
    \int_0^{L_e}r_e(x)^*z_e(x)\dd x .
    \label{eq:space}
\end{equation}
Define
\begin{equation}
\begin{aligned}
&D(\cA)
:=
\Bigl\{
r\in\prod_{e=1}^N H^1(0,L_e;\C^4):
\eqref{eq:linear-boundary}\text{ holds at all }\junction_e
\Bigr\},
\\
&(\cA r)_e=-A_er'_e+J_er_e .
\end{aligned}
\label{eq:operator-A}
\end{equation}
Integration by parts gives
\begin{equation}
    (\cA^*z)_e=A_e^*z'_e+J_e^*z_e,
\end{equation}
with 
\begin{equation}
D(\cA^*)
=
\left\{
z\in\prod_{e=1}^N H^1(0,L_e;\C^4):
\eqref{eq:adjoint-boundary}\text{ holds}
\right\},
\label{eq:adjoint-domain}
\end{equation}
where, for all $e\in\mathcal E$,
\begin{equation}
\begin{aligned}
(z_e(L_e))_j
&=p_{\junction_e}(z_{e+1}(0))_j,
&& j\in\{1,3\},\\
(z_{e+1}(0))_i
&=p_{\junction_e}(z_e(L_e))_i,
&& i\in\{2,4\},
\end{aligned}
\label{eq:adjoint-boundary}
\end{equation}

Let
$
    d_u:=2\sum_{e=1}^N N_e .
$
The complexified bounded control operator
$\cB:\C^{d_u}\to\cH$ is defined by
\begin{equation}
    (\cB U)_e
    =
    \sum_{j=1}^{N_e}
    \mathbf1_{\omega_{e,j}}B_eu_{e,j},
\end{equation}
and its adjoint is
\begin{equation}
    (\cB^*z)_{e,j}
    =
    \int_{\omega_{e,j}}B_e^*z_e(x)\dd x .
\end{equation}
Thus the linearized system
\eqref{eq:linear-physical},~\eqref{eq:linear-boundary} takes the
abstract form
\begin{equation}
    \partial_t r(t)
    =
    \cA r(t)+\cB U(t),
    \label{eq:abstract_linearized}
\end{equation}
where
$U(t)\in\R^{d_u}$ collects all localized guidance inputs
$u_{e,j}$.

Under Assumption~\ref{ass:model}, using~\eqref{eq:det-Aek}, each $A_e$ is diagonalizable with
two positive and two negative real characteristic speeds. Moreover, the junction conditions~\eqref{eq:linear-boundary} uniquely determine the incoming characteristic traces from the outgoing ones. Therefore, the linearized system's~\eqref{eq:linear-physical},~\eqref{eq:linear-boundary}
well-posedness in $\cH$ follows
from~\cite[Appendix~A, Theorem~A.4]{bastin_coron}. The well-posedness in $H^1$ follows from~\cite[Appendix~A, Theorem~A.1]{bastin_coron}.

The actuator-placement question is recast as a spectral detectability problem for the abstract linearized system \eqref{eq:abstract_linearized}. More precisely, for a prescribed decay rate $\decay>0$, we seek actuator configurations such that every adjoint eigenmode associated with an eigenvalue satisfying $\Re s\geq -\decay$ is detected by the localized guidance input. This modal detectability property is characterized by the Fattorini-Hautus condition~\cite{fattorini1966some}, namely
\begin{equation}
    \ker(sI-\cA^*)\cap\ker\cB^*
    =
    \{0\},
    \quad
    \forall s\in \C, \Re s\geq-\decay .
   \tag{FH}\label{eq:FH}
\end{equation}
Indeed, any adjoint eigenmode in $\ker\cB^*$ is unaffected by bounded
feedback and therefore cannot be controlled.
\section{Main Result}
\label{sec:main_result}
In the following, we prove that~\eqref{eq:FH} can be satisfied with only one corridor actuated with two guidance patches (i.e., zones with actuator biasing lane-changes in both directions).
For $\decay>0$, define the set of eigenvalues
\begin{equation}
    \mathcal S_{\decay}
    :=
    \left\{
        s\in\C:\;
        \Re s\geq-\decay,\;
        \Ker(sI-\cA^*)\neq\{0\}
    \right\}.
    \label{eq:Sdelta}
\end{equation}
\begin{theorem}[Two patches are sufficient for~\eqref{eq:FH}]
    \label{thm:main}
   Suppose Assumptions~\ref{ass:model}
and~\ref{ass:transport-nondegeneracy} hold. Let $e_\star$ be as in Assumption~\ref{ass:transport-nondegeneracy} $d_c>0$, and
    fix
$a\in(0,L_{e_\star})$. 
Then, there exist $\delta>0$ and $\ell>0$, with
$\ell<\delta$ and $a+\delta+\ell<L_{e_\star}$, such that choosing in~\eqref{eq:indicator-control}
\begin{equation}
 \omega_{e_\star, 1}=(a,a+\ell),\quad
 \omega_{e_\star, 2}=(a+\delta,a+\delta+\ell)
 \label{eq:two-patches-selected-edge},
\end{equation}
the criterion~\eqref{eq:FH} is satisfied.
Moreover, among all pairs \((\delta,\ell)\) satisfying \(0<\ell<\delta\) and \(a+\delta+\ell<L_{e_\star}\), the set of choices for which~\eqref{eq:FH} holds is residual and has full Lebesgue measure.
In particular, the admissible patch geometries satisfying~\eqref{eq:FH} form a dense set of full Lebesgue measure.
\end{theorem}
\begin{proof}
By~\eqref{eq:kappa-nonzero-guidance},
$
    \Ran B_{e_\star}=\mathcal D.
$
Together with~\eqref{eq:range-Je}, this implies that there exists
$Q_{e_\star}\in\C^{2\times2}$ such that
\begin{equation}
    J_{e_\star}A_{e_\star}^{-1}B_{e_\star}
    =
    B_{e_\star}Q_{e_\star}.
    \label{eq:JAinvB}
\end{equation}

We first establish a local rank property on the actuated corridor for
nonzero eigenvalues. Writing
$
    K_{e_\star}
    :=
    \diag(\kappa_{e_\star,+},\kappa_{e_\star,-}),
$
one has
\begin{equation}
    B_{e_\star}
    =
    \begin{pmatrix}
        -K_{e_\star}\\
        K_{e_\star}
    \end{pmatrix},
    \quad
    A_{e_\star}^{-1}B_{e_\star}
    =
    \begin{pmatrix}
        -A_{e_\star,1}^{-1}K_{e_\star}\\
        A_{e_\star,2}^{-1}K_{e_\star}
    \end{pmatrix}.
\end{equation}
Adding the upper block-row to the lower block-row gives
\begin{align}
 &\det
 \bigl[
 B_{e_\star},
 A_{e_\star}^{-1}B_{e_\star}
 \bigr]
 \notag\\
 &\quad=
 (\kappa_{e_\star,+}\kappa_{e_\star,-})^2
 \det\!\left(
 A_{e_\star,2}^{-1}-A_{e_\star,1}^{-1}
 \right)
 \notag\\
 &\quad=
 -\frac{
 2(\kappa_{e_\star,+}\kappa_{e_\star,-})^2
 \left(
 M_{e_\star,1}\bar{\totald}_{e_\star,1}
 -
 M_{e_\star,2}\bar{\totald}_{e_\star,2}
 \right)^2
 }{
 \det A_{e_\star,1}\det A_{e_\star,2}
 },
 \label{eq:local-rank-main-proof}
\end{align}
where we used
$
A_{e_\star,2}^{-1}-A_{e_\star,1}^{-1}
=
A_{e_\star,2}^{-1}
(A_{e_\star,1}-A_{e_\star,2})
A_{e_\star,1}^{-1}.
$
By Assumption~\ref{ass:transport-nondegeneracy},
\begin{equation}
    \det
    \bigl[
    B_{e_\star},
    A_{e_\star}^{-1}B_{e_\star}
    \bigr]
    \neq0.
    \label{eq:BAinvB-nonzero}
\end{equation}

Let now
$
    s\in\mathcal S_{\decay}\setminus\{0\}.$
On corridor $e_\star$, an adjoint eigenfunction satisfies
\begin{equation}
    z'_{e_\star}(x)
    =
    \mathsf M_{e_\star,s}z_{e_\star}(x),
    \quad
    \mathsf M_{e_\star,s}
    :=
    (A_{e_\star}^*)^{-1}
    (sI-J_{e_\star}^*).
    \label{eq:Ms-main-proof}
\end{equation}
Using~\eqref{eq:JAinvB},
\begin{equation}
\begin{aligned}
&
\bigl[
B_{e_\star},
\mathsf M_{e_\star,s}^*B_{e_\star}
\bigr]
\\
&\quad=
\bigl[
B_{e_\star},
(\bar s I-J_{e_\star})
A_{e_\star}^{-1}B_{e_\star}
\bigr]
\\
&\quad=
\bigl[
B_{e_\star},
\bar s A_{e_\star}^{-1}B_{e_\star}
-
B_{e_\star}Q_{e_\star}
\bigr].
\end{aligned}
\end{equation}
Adding the first block column multiplied by $Q_{e_\star}$ to the
second one and using~\eqref{eq:BAinvB-nonzero} gives
\begin{equation}
    \det
    \bigl[
    B_{e_\star},
    \mathsf M_{e_\star,s}^*B_{e_\star}
    \bigr]
    =
    \bar s^{\,2}
    \det
    \bigl[
    B_{e_\star},
    A_{e_\star}^{-1}B_{e_\star}
    \bigr]
    \neq0.
    \label{eq:local-rank-Ms}
\end{equation}
Equivalently,
\begin{equation}
    \rank
    \begin{pmatrix}
        B_{e_\star}^*\\
        B_{e_\star}^*\mathsf M_{e_\star,s}
    \end{pmatrix}
    =4.
    \label{eq:local-observability-rank}
\end{equation}

For $\delta>0$, define
\begin{equation}
    \mathcal O_{e_\star,s}(\delta)
    :=
    \begin{pmatrix}
        B_{e_\star}^*\\
        B_{e_\star}^*
        \e^{\mathsf M_{e_\star,s}\delta}
    \end{pmatrix}.
\end{equation}
Subtracting the first block row from the second and dividing the latter
by $\delta$ yields
\begin{equation}
    \begin{pmatrix}
        B_{e_\star}^*\\[1mm]
        \displaystyle
        \frac{
        B_{e_\star}^*
        (\e^{\mathsf M_{e_\star,s}\delta}-I)
        }{\delta}
    \end{pmatrix}
    \longrightarrow
    \begin{pmatrix}
        B_{e_\star}^*\\
        B_{e_\star}^*\mathsf M_{e_\star,s}
    \end{pmatrix}
    ~
    \rm as~\delta\to0^+.
\end{equation}
Thus~\eqref{eq:local-observability-rank} implies that, for each fixed
$s\in\mathcal S_{\decay}\setminus\{0\}$,
\begin{equation}
    \rank\mathcal O_{e_\star,s}(\delta)=4,
\end{equation}
for all sufficiently small $\delta>0$.

We now pass from point observations to the two guidance patches. For
$
    0<\ell<\delta,~
    a+\delta+\ell<L_{e_\star},
$
define
\begin{equation}
    \mathcal G_{e_\star,s}(\delta,\ell)
    :=
    \begin{pmatrix}
    \displaystyle
    \int_0^\ell
        B_{e_\star}^*
        \e^{\mathsf M_{e_\star,s}\xi}
        \dd\xi
    \\[4mm]
    \displaystyle
    \int_0^\ell
        B_{e_\star}^*
        \e^{\mathsf M_{e_\star,s}(\delta+\xi)}
        \dd\xi
    \end{pmatrix}.
    \label{eq:Gs-main-proof}
\end{equation}
For fixed $\delta$,
\begin{equation}
    \frac1\ell
    \mathcal G_{e_\star,s}(\delta,\ell)
    \longrightarrow
    \mathcal O_{e_\star,s}(\delta)
    ~ \rm as~
    \ell\to0^+.
\end{equation}
Consequently, for all fixed
$s\in\mathcal S_{\decay}\setminus\{0\}$ there exists at least one
admissible pair $(\delta,\ell)$ for which
\begin{equation}
    \det
    \mathcal G_{e_\star,s}(\delta,\ell)
    \neq0.
    \label{eq:Gs-nontrivial}
\end{equation}

We next treat separately the possible eigenvalue $s=0$. Suppose that
$z\in\Ker\cA^*$. On corridor $e_\star$,
\begin{equation} \label{eq:zero_eigenfunction}
    z'_{e_\star}
    =
    -(A_{e_\star}^*)^{-1}
    J_{e_\star}^*z_{e_\star}.
\end{equation}
Set, for all $x\in [0, L_{e_\star}]$,
$
    y(x):=B_{e_\star}^*z_{e_\star}(x).
$
Using~\eqref{eq:JAinvB}, we obtain
\begin{equation}\begin{aligned}
    y'(x)
    &=
    -B_{e_\star}^*
    (A_{e_\star}^*)^{-1}
    J_{e_\star}^*z_{e_\star}(x)
    \\
    &=
    -\bigl(
    J_{e_\star}A_{e_\star}^{-1}B_{e_\star}
    \bigr)^*
    z_{e_\star}(x)
    \\
    &=
    -Q_{e_\star}^*y(x).
\end{aligned} \end{equation}
Therefore, if $a+\xi\in [0, L_{e_\star}]$,
\begin{equation}
    y(a+\xi)
    =
    \e^{-Q_{e_\star}^*\xi}y(a).
    \label{eq:zero-output-evolution}
\end{equation}
Define
\begin{equation}
    \mathcal R_0(\ell)
    :=
    \int_0^\ell
    \e^{-Q_{e_\star}^*\xi}\dd\xi .
    \label{eq:R0-zero-mode}
\end{equation}
Since
\begin{equation} \label{eq:R0_invertible}
    \frac1\ell\mathcal R_0(\ell)
    \longrightarrow I_2
    \quad
    \rm as ~\ell\to0^+,
\end{equation}
$\mathcal R_0(\ell)$ is invertible for all sufficiently small
$\ell>0$.

We now choose one pair $(\delta_ ,\ell)$  simultaneously for all
eigenvalues. Set
\begin{equation}
    \Omega_a
    :=
    \left\{
        (\delta,\ell)\in\R^2:
        0<\ell<\delta,\;
        a+\delta+\ell<L_{e_\star}
    \right\}.
\end{equation}
For
$s\in\mathcal S_{\decay}\setminus\{0\}$, define the Grammian determinant
\begin{equation}
    \Psi_s(\delta,\ell)
    :=
    \det\!\left(
        \mathcal G_{e_\star,s}(\delta,\ell)^*
        \mathcal G_{e_\star,s}(\delta,\ell)
    \right).
    \label{eq:Psi-main-proof}
\end{equation}
The function $\Psi_s$ is non-negative, real valued and analytic on the connected open set
$\Omega_a$, and
\begin{equation}
    \Psi_s(\delta,\ell)>0
    \quad\Longleftrightarrow\quad
    \rank
    \mathcal G_{e_\star,s}(\delta,\ell)=4.
\end{equation}
By~\eqref{eq:Gs-nontrivial}, $\Psi_s$ is not identically zero.
Therefore, by the zero set theorem for nontrivial real-analytic
functions~\cite{mityagin2020zero},
\begin{equation}
    Z_s
    :=
    \left\{
        (\delta,\ell)\in\Omega_a:
        \Psi_s(\delta,\ell)=0
    \right\},
\end{equation}
has Lebesgue measure zero. Since $Z_s$ is relatively closed in
$\Omega_a$, it is nowhere dense.

For the possible zero eigenvalue, define
\begin{equation}
    \Psi_0(\delta,\ell)
    :=
    \det\!\left(
        \mathcal R_0(\ell)^*
        \mathcal R_0(\ell)
    \right).
    \label{eq:Psi-zero-mode}
\end{equation}
This is also non-negative, real valued, analytic on $\Omega_a$ and not identically zero because of~\eqref{eq:R0_invertible}. Hence by~\cite{mityagin2020zero},
\begin{equation}
    Z_0
    :=
    \left\{
        (\delta,\ell)\in\Omega_a:
        \Psi_0(\delta,\ell)=0
    \right\},
    \label{eq:Z-zero-mode}
\end{equation}
has Lebesgue measure zero and is nowhere dense in $\Omega_a$.

By Proposition~\ref{prop:test_zero_char} in Appendix~\ref{appendix:prop_countable},
$\mathcal S_{\decay}$ is at most countable. It follows that
\begin{equation}
    Z
    :=
    Z_0
    \cup
    \bigcup_{s\in\mathcal S_{\decay}\setminus\{0\}}Z_s,
\end{equation}
is a countable union of nowhere dense sets and of sets of Lebesgue
measure zero. Thus
\begin{equation}
    \Omega_a^{\rm valid}
    :=
    \Omega_a\setminus Z,
\end{equation}
is residual and has full Lebesgue measure in $\Omega_a$. In particular,
it is dense and nonempty.

Choose now
$(\delta,\ell)\in\Omega_a^{\rm valid}$ and let
\begin{equation} \label{eq:eigenfunction}
    z\in
    \Ker(sI-\cA^*)
    \cap
    \Ker\cB^*,
    ~ \rm{for}~s\in \C~\rm{such~ that}~
    \Re s\ge-\decay.
\end{equation}
If $s\notin\mathcal S_{\decay}$, then $z=0$ by definition. Therefore, $s\in\mathcal S_{\decay}$.

First assume $s\neq0$,so that~\eqref{eq:eigenfunction} gives on corridor $e_\star$
\begin{equation}
    \int_{\omega_{e_\star, 1}}
    B_{e_\star}^*z_{e_\star}(x)\dd x=0,
    \quad
    \int_{\omega_{e_\star, 2}}
    B_{e_\star}^*z_{e_\star}(x)\dd x=0.
    \label{eq:vanishing}
\end{equation}
Using~\eqref{eq:Ms-main-proof}, if $a+\xi\in [0, L_{e_\star}]$
\begin{equation}
    z_{e_\star}(a+\xi)
    =
    \e^{\mathsf M_{e_\star,s}\xi}
    z_{e_\star}(a),
\end{equation}
and hence~\eqref{eq:vanishing} is equivalent to
\begin{equation}
    \mathcal G_{e_\star,s}(\delta,\ell)
    z_{e_\star}(a)=0.
\end{equation}
Since $(\delta,\ell)\notin Z_s$,
$\mathcal G_{e_\star,s}(\delta,\ell)$ is invertible, and therefore
$
    z_{e_\star}(a)=0.
$
Uniqueness for~\eqref{eq:Ms-main-proof} yields
$z_{e_\star}\equiv0$. The adjoint junction conditions
\eqref{eq:adjoint-boundary}, together with
$p_{\junction_e}>0$, then propagate this vanishing through the cycle,
and consequently $z\equiv0$.

It remains to consider $s=0$, by~\eqref{eq:eigenfunction}, the first patch gives
\begin{equation}
    0
    =
    \int_a^{a+\ell}
    B_{e_\star}^*z_{e_\star}(x)\dd x.
\end{equation}
By~\eqref{eq:zero-output-evolution},
\begin{equation}
    \mathcal R_0(\ell)y(a)=0.
\end{equation}
Since $(\delta,\ell)\notin Z_0$,
$\mathcal R_0(\ell)$ is invertible, and therefore
$
    y(a)=0.$
Equation~\eqref{eq:zero-output-evolution} then gives
\begin{equation}
    B_{e_\star}^*z_{e_\star}(x)=0,
    \quad
    \forall x\in [0,  L_{e_\star}].
\end{equation}
By~\eqref{eq:kappa-nonzero-guidance} and~\eqref{eq:range-Je}, 
$
    \Ran J_{e_\star}
    \subset
    \Ran B_{e_\star},$ hence,
$
    \Ker B_{e_\star}^*
    \subset
    \Ker J_{e_\star}^*.$
Thus
\begin{equation}
    J_{e_\star}^*z_{e_\star}(x)=0,
\end{equation}
and~\eqref{eq:zero_eigenfunction} yields, for all $x\in [0, L_{e_\star}]$
$
    z'_{e_\star}(x)=0.
$
Consequently, 
\begin{equation}
    z_{e_\star}
    \equiv \col(
        c_+, c_-, c_+, c_-)
    \label{eq:zero_mode_constant},
\end{equation}
for some $c_+,c_-\in\C$.
Here we used the fact that, since $\rank B_{e_\star}=2$, $\dim\Ker B_{e_\star}^*=2$. Moreover,
$
\col(c_+,c_-,c_+,c_-)\in\Ker B_{e_\star}^*
$
for all $c_+,c_-\in\C$. Hence
\begin{equation}
\Ker B_{e_\star}^*
=
\left\{
\col(c_+,c_-,c_+,c_-):c_+,c_-\in\C
\right\}.
\end{equation}

Then at $\junction_e$, injecting~\eqref{eq:zero_mode_constant} in~\eqref{eq:adjoint-boundary} gives
\begin{equation}\label{eq:zeplus1}
    z_{e+1}(0)
    =
    \col(
        p_{\junction_e}^{-1}c_+,
        p_{\junction_e}c_-,
        p_{\junction_e}^{-1}c_+,
        p_{\junction_e}c_-)
    \in\Ker B_{e+1}^*.
\end{equation}
Since
$
    \Ker B_{e+1}^*
    \subset
    \Ker J_{e+1}^*,
$
the constant function equal to $z_{e+1}(0)$~\eqref{eq:zeplus1} is an eigenfunction associated to the eigenvalue $s=0$ in corridor $e+1$. By uniqueness,
$z_{e+1}$ is constant in that corridor. Iterating around the cycle and
returning to corridor $e_\star$ gives
\begin{equation}
    c_+
    =
    P_{\rm cyc}^{-1}c_+,
    \quad
    c_-
    =
    P_{\rm cyc}c_-.
\end{equation}
By Assumption~\ref{ass:model} and by definition,
$0<P_{\rm cyc}<1$. It follows that
$
    c_+=c_-=0.
$
Hence $z_{e_\star}\equiv0$, and the above propagation through the cyclic corridor
gives $z\equiv0$.

We conclude that~\eqref{eq:FH} holds.
\end{proof}
\FloatBarrier
\section{Numerical Example}
\label{section:numerics}
We present a numerical example in this section to illustrate the proof of Theorem~\ref{thm:main}.
We consider~\eqref{eq:linear-physical},~\eqref{eq:linear-boundary} with $N=5$ corridors of lengths
$
(L_1,L_2,L_3,L_4,L_5)=(10,11,9,10.5,9.5).
$
We choose for all $e\in \mathcal E$,
$
V=1.5,~ \rho^\star=6,~
M_{e,1}=0.2,~ M_{e,2}=0.3,~
\lambda_0=0.35,~ \theta=1.$
The same external arrival flux is imposed at every junction,
\begin{equation}
\bar g_{J_e}
=
\col(
0.50,
0.01,
0.01,
0.50
),
\end{equation}
with ordering $(1,+),(1,-),(2,+),(2,-)$, which induces a strong directional segregation between the two lanes (see remark before Assumption~\ref{ass:transport-nondegeneracy}). The continuation ratios are given by
$
p_{J_e}=0.67,~\forall e\in \mathcal E.
$
The decay rate is
$
d_c=0.1$.
The spectrum is approximated by Chebyshev collocation.
The actuated corridor is selected according to (see~\eqref{eq:cond_corridor_non_degen})
\begin{equation}
e^\star\in
\operatorname*{arg\,max}_{e\in\mathcal E}
\left|
M_{e,1}\bar R_{e,1}
-
M_{e,2}\bar R_{e,2}
\right|,
\end{equation}
and the first guidance patch starts at
$
a=0.15L_{e^\star}.$
\begin{figure}[tb]
    \centering
    \includegraphics[width=0.8\linewidth]{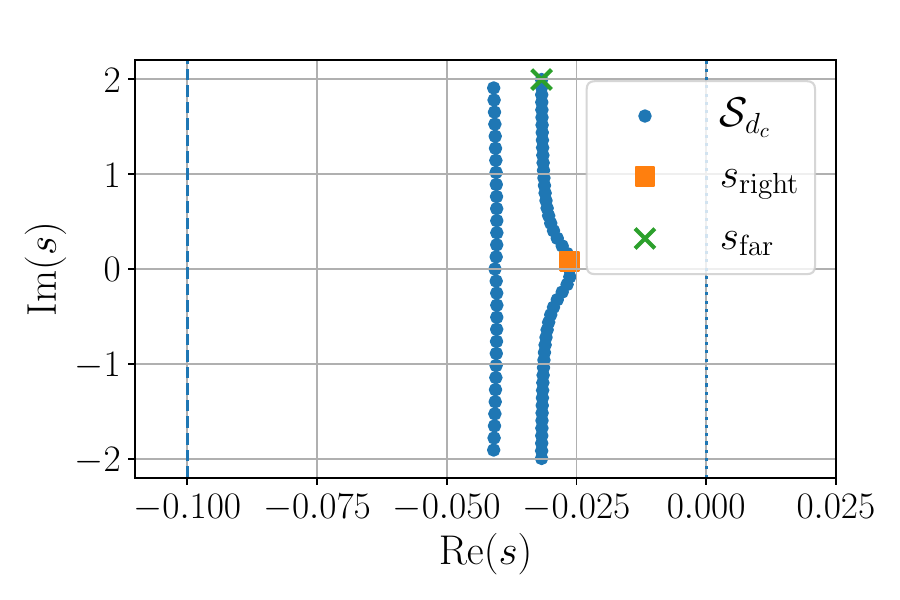}
    \caption{Critical spectrum $\mathcal S_{d_c}$~\eqref{eq:Sdelta} intersected with a rectangle.}
    \label{fig:spectrum-free}
\end{figure}
\begin{figure}[tb]
    \centering
    \begin{minipage}{0.7\linewidth}
        \centering
        \includegraphics[width=\linewidth]{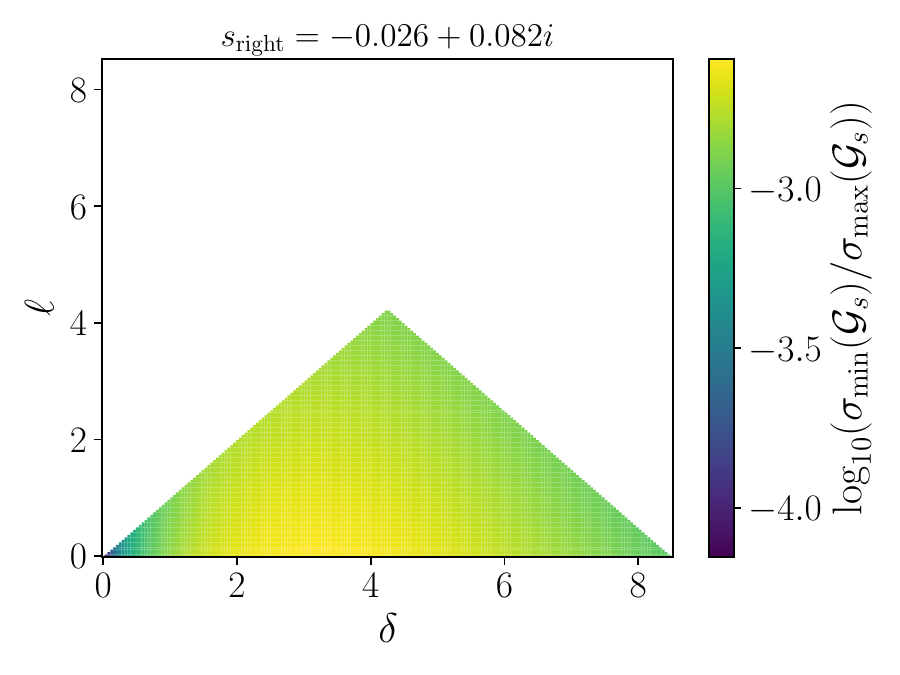}
    \end{minipage}
    \hfill
    \begin{minipage}{0.7\linewidth}
        \centering
        \includegraphics[width=\linewidth]{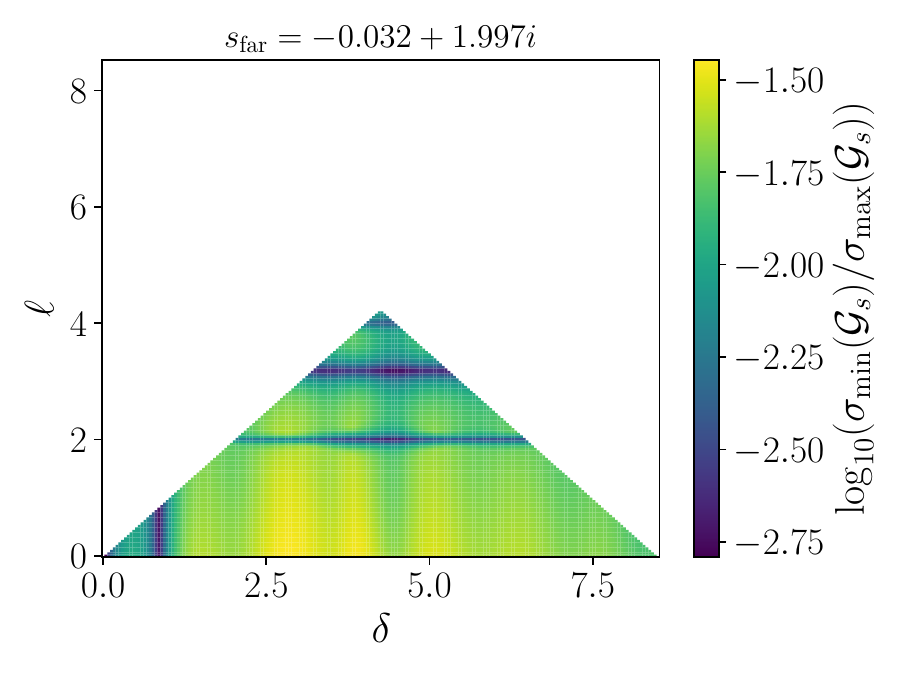}
    \end{minipage}
    \caption{Relative observability margin
    as a function of the patch geometry
    for two eigenvalues.}
    \label{fig:conditioning-free}
\end{figure}
\begin{figure}[tb]
    \centering
    \includegraphics[width=0.7\linewidth]{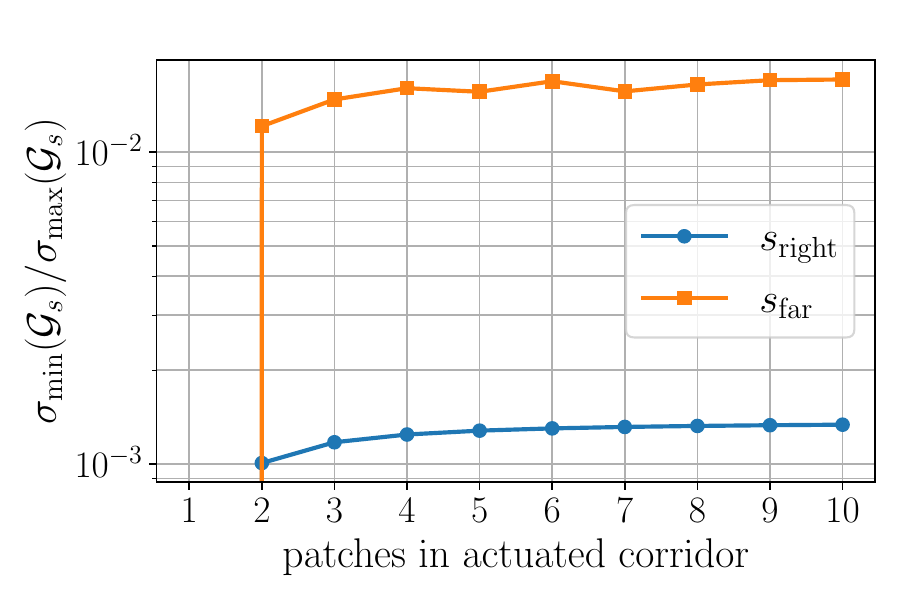}
    \caption{Relative observability margin for uniformly distributed patches
in a single actuated corridor, with total guided length fixed to one.}
    \label{fig:multi_patch}
\end{figure}
Fig.~\ref{fig:spectrum-free} shows that all critical eigenvalues lie in the left half-plane. This stability property was consistently observed numerically for the free-flow configurations considered here. However, most eigenvalues remain relatively close to the imaginary axis, indicating the presence of slowly decaying modes.
The conditioning of the modal observability test~\eqref{eq:Gs-main-proof} in Fig.~\ref{fig:conditioning-free} is poorer for eigenvalues close to the origin. This is consistent with~\eqref{eq:local-rank-Ms}, since the local determinant becomes small as $s\to 0$. In Fig.~\ref{fig:conditioning-free}, the rightest eigenvalue
$
s_{\mathrm{right}}\approx -0.026+0.082\,i
$
is consequently much more poorly conditioned than the higher-frequency mode
$
s_{\mathrm{far}}\approx -0.032+1.997\,i.
$
If $s=0$ were an  eigenvalue,~\eqref{eq:R0_invertible} would ensure observability.
The poorly conditioned patch geometries appear as thin curves in the $(\delta,\ell)$-plane for $s_{\rm far}$. This is consistent with the exceptional zero Lebesgue measure zero sets arising in the proof of Theorem~\ref{thm:main}. Admissible geometries form a full measure set.

To further study the effect of distributing the actuation, we fix the total
guided length in the actuated corridor to one and split it into $n\in \{1,\ldots, 10\}$ uniformly
distributed patches. Fig.~\ref{fig:multi_patch} shows the obtained relative
observability margin. For $n=1$, the observation matrix cannot have full rank, so the observability test fails. For $n\geq2$, the margin becomes positive,
while increasing the number of patches leads only to a small improvement in
conditioning. This is expected since the total guided length is fixed, so
adding patches only redistributes the same amount of actuation (biasing lane-changes) over the corridor.

\section{Conclusions and Future Works} \label{sec:conclusion}
We have established physically meaningful sufficient conditions on the model parameters and on the geometry of localized guidance zones that guarantee the spectral detectability condition (FH) for a bidirectional flow pedestrian model on a cyclic corridor. Natural directions for future work include the extension of the analysis to congested and mixed free-flow/congested regimes, and the verification of an additional robustness assumption required in~\cite{braun2026stabilization} to design a stabilizing feedback law as in~\cite{Jean_Adam_Ouidir_Mathieu, braun_corona}.




\appendix
\setcounter{subsection}{0}
\renewcommand{\thesubsection}{\Alph{subsection}}
\renewcommand{\thetheorem}{\Alph{subsection}.\arabic{theorem}}
\subsection{Proof of Lemma~\ref{lem:inflow_gives_equi}}
\label{appendix:proof_lem_{e+1}nflow_equi}
    Iterating once around the concourse gives for the positive direction part of~\eqref{eq:boundary_equi}
\begin{equation}\label{eq:equilibrium-flux-from-inflow+}
    (1-P_{\rm cyc})\bar f_{e,k,+}
    =
    \Gamma_{e,k}^{+}.
\end{equation}
Similarly,
the negative direction part of~\eqref{eq:boundary_equi} becomes
\begin{equation}\label{eq:equilibrium-flux-from-inflow-}
    -(1-P_{\rm cyc})\bar f_{e,k,-}
    =
    \Gamma_{e,k}^{-}.
\end{equation}
Hence, under~\eqref{eq:equilibrium-cycle-loss} the stationary fluxes are uniquely determined by the external arrival.
In particular,
\begin{equation}
    \bar f_{e,k,+}-\bar f_{e,k,-}
    =
    \frac{
        \Gamma_{e,k}^{+}+\Gamma_{e,k}^{-}
    }{1-P_{\rm cyc}}
    >0.
    \label{eq:equilibrium-total-flux}
\end{equation}
At equilibrium,
\begin{equation}
    Q_{e,k}:=\bar f_{e,k,+}-\bar f_{e,k,-}
    =
    \bar\totald_{e,k}
    \left(
        V-M_{e,k}\bar\totald_{e,k}
    \right).
    \label{eq:equilibrium-fundamental-diagram}
\end{equation}
By~\eqref{eq:inflow-admissibility}, this equation has a unique solution
on the interval
\begin{equation}
    0<\totald_{e,k}
    <
    \frac{V}{2M_{e,k}},
\end{equation}
namely
\begin{equation}
    \bar{\totald}_{e,k}
    =
    \frac{
        V-\sqrt{V^2-4M_{e,k}Q_{e,k}}
    }{2M_{e,k}}.
    \label{eq:equilibrium-total-density}
\end{equation}
Since
$
    V-M_{e,k}\bar{\totald}_{e,k}>0,
$
we have
\begin{equation}
    \bar\rho_{e,k,+}
    =
    \frac{\bar f_{e,k,+}}
    {V-M_{e,k}\bar{\totald}_{e,k}},
    \quad
    \bar\rho_{e,k,-}
    =
    \frac{-\bar f_{e,k,-}}
    {V-M_{e,k}\bar{\totald}_{e,k}}.
    \label{eq:equilibrium-directional-densities}
\end{equation}
Conditions~\eqref{eq:positive-effective-arrivals} therefore imply
\begin{equation}
    \bar\rho_{e,k,+}>0,
    \quad
    \bar\rho_{e,k,-}>0,
\end{equation}
and, by construction,
\begin{equation}
    \bar\rho_{e,k,+}
    +
    \bar\rho_{e,k,-}
    =
    \bar{\totald}_{e,k}.
\end{equation}
Thus, all constant densities are completely
determined by $\bar g$.
It remains to enforce the source equilibrium
\eqref{eq:source_zero} by choosing $\bar \eta$.  For $\lambda_0>0$, the common factor
$\lambda_0$ cancels from this relation.  Using~\eqref{eq:rates} and 
\begin{equation}
    \frac{\sigma(d)}{\sigma(-d)}=\e^d,
\end{equation}
equation~\eqref{eq:source_zero} is equivalent to
\begin{equation}
    \exp(\bar d_{e,\alpha})
    =
    \frac{\bar\rho_{e,2,\alpha}}
         {\bar\rho_{e,1,\alpha}}
    \frac{\chi(\bar{\totald}_{e,1})
    }{
        \chi(\bar{\totald}_{e,2})
    },
    \label{eq:equilibrium-d-from-density}
\end{equation}
with
\begin{equation}
    \bar d_{e,\alpha}
    =
    \frac{
        M_{e,1}\bar{\totald}_{e,1}
        -
        M_{e,2}\bar{\totald}_{e,2}
        +
        \bar\eta_{e,\alpha}
    }{\theta}.
\end{equation}
The nominal guidance required to satisfy~\eqref{eq:source_zero} is then uniquely determined by
\begin{equation}
    \bar\eta_{e,\alpha}
    =
    \theta
    \log\!\left(
        \frac{\bar\rho_{e,2,\alpha}}
             {\bar\rho_{e,1,\alpha}}
        \frac{
            1-\bar{\totald}_{e,1}/\rho^\star
        }{
            1-\bar{\totald}_{e,2}/\rho^\star
        }
    \right)
    -
    \left(
        M_{e,1}\bar{\totald}_{e,1}
        -
        M_{e,2}\bar{\totald}_{e,2}
    \right).
\end{equation}
Consequently, under Assumption~\ref{ass:model},
$\bar g$ uniquely determines an equilibrium point. \hfill$\blacksquare$
\subsection{Eigenvalues are countable}
\label{appendix:prop_countable}
\begin{proposition}
\label{prop:test_zero_char}
For all $\decay>0$, and all $C>0$,
\begin{equation}
    \mathcal S_{\decay}\cap\{s\in\C:\ |s|\le C\},
\end{equation}
is finite. In particular, $\mathcal S_{\decay}$ is at most countable. 
\end{proposition}
\begin{proof}
Since each $A_e$ is invertible, using~\eqref{eq:operator-A} we have for $r\in D(\cA)$
\begin{equation}
    r'_e
    =
    A_e^{-1}J_er_e
    -
    A_e^{-1}(\cA r)_e .
\end{equation}
Hence, for some constant $C>0$,
\begin{equation}
    \|r\|_{\prod_e H^1(0,L_e;\C^4)}
    \le
    C\bigl(
        \|r\|_{\cH}+\|\cA r\|_{\cH}
    \bigr).
\end{equation}
Thus $D(\cA)$, endowed with the graph norm, is continuously embedded
in $\prod_e H^1(0,L_e;\C^4)$. Since the number of corridors is finite
and
\begin{equation}
    H^1(0,L_e;\C^4)
    \hookrightarrow
    L^2(0,L_e;\C^4),
\end{equation}
is compact, see, e.g.,~\cite[Thm.~8.8]{brezisfunctionnal}, the
embedding
\begin{equation}
    D(\cA)\hookrightarrow\cH,
\end{equation}
is compact. Finally, since $\cA$ generates a $C_0$-semigroup (because the linearized system is well-posed),
$\rho(\cA)\neq\varnothing$. Hence, by~\cite[Prop.~II.4.25]{engel-nagel}, $\cA$ has compact
resolvent. Therefore, by~\cite[Ch.~III, \S6.8, Thm.~6.29]{kato},
$\sigma(\cA)$ consists of isolated eigenvalues of finite algebraic
multiplicity, with no finite accumulation point.

Moreover,
\begin{equation}
(\lambda I-\mathcal A^*)^{-1}
=
\big((\overline{\lambda}I-\mathcal A)^{-1}\big)^*,
\quad
\forall \lambda\in\rho(A^*),
\end{equation}
so $\cA^*$ also has compact resolvent and hence has the same
spectral properties.
\end{proof}




\bibliographystyle{abbrv}
\bibliography{references}

\end{document}